\documentclass[11pt,a4paper]{article}
\usepackage{amssymb}
\usepackage{amsmath}
\usepackage{graphics}
\usepackage{epsfig}
\usepackage{float}
\usepackage{graphicx}
\usepackage{amsfonts}
\usepackage{amsthm}
\usepackage{newlfont}
\usepackage{color}
\usepackage{subcaption}
\usepackage[numbers]{natbib}

\begin{document}

\title{Energy eigen values of quadratic, pure quartic and quartic anharmonic oscillators with variational method}
\author{Shaheen Irfan$^{a}$, Zaki Ahmad$^{b,d}$ \thanks{zaki.ahmad@gcwus.edu.pk, zakiahmad754@gmail.com}, Nosheen Akbar$^{c}$\thanks{nosheenakbar@cuilahore.edu.pk}, Minal Mansoor$^{d}$, Hussnain Sumbul$^{b}$ \\
$^{a}$\textit {Department of Physics, The University of Lahore, Lahore, Pakistan.}\\
$^{b}$\textit {Center for High Energy Physics, Punjab University, Lahore, Pakistan.}\\
$^{c}$\textit {COMSATS University Islamabad, Lahore Campus, Pakistan.}\\
$^{d}$\textit {Govt. College Women University, Sialkot, Pakistan.}}
\maketitle
\begin{abstract}
In this work, the energy eigen values are calculated for the quadratic ($\frac{g^2 x^2}{2}$), pure quartic ($\lambda x^4 $), and quartic anharmonic oscillators ($\frac{g^2 x^2}{2} + \lambda x^4 $) by applying variational method. For this, simple harmonic oscillator wave functions are considered as trial wave functions to calculate the energies for the ground state and first ten excited states with $g = 1$ and $\lambda =1/4$. For quartic anharmonic oscillators, energy values are calculated at different values of $\lambda$ with $g=1$. These energies for the ground state are compared with available numerically calculated data. Maximum value of $\%$error is found to be 1.9977. To get more accurate results, a new set of trial wave functions is suggested. With the newly proposed wave functions, maximum value of $\%$ error for the energy values reduces to 0.561. In this work, energies for the ground and first five excited states of quartic anharmonic oscillators are reported at different values of $\lambda$. Dependence of $\lambda$ on the wave functions is observed and concluded that wave functions are converging (shrinking) by increasing the $\lambda$.

\end{abstract}

\bigskip

\section{\protect\bigskip \textbf{Introduction}}
The fundamental potential model ($\frac{g^2 x^2}{2} + \lambda x^4 $) with non-zero values of $g$ and $\lambda$ deviates from harmonic oscillator so named it as quartic anharmonic oscillator (QAO) potential. In this potential model $``x"$ is the representation of displacement from equilibrium position. QAO potential has many applications in particle physics, quantum chemistry, quantum field theory, laser theory and other areas. A significant application of this model is to understand the phenomena of different quantum mechanical systems with vibrational degrees of freedom. QAO model can also be used to describe the crystal lattice vibrations in solid-state theory.

Different techniques like  WKB \cite{5,6}, phase-integral method (PIM) based on the generalized Bohr-Sommerfeld quantization \cite{7}, variational approach based on wavefunctions (WFs) parameterization \cite{8}, and variational method based on path integrals \cite{9,10}
have been used to investigate QAO. In ref. \cite{Adali}, energy eigen values and wavefunctions for QAO are calculated with Perturbation theory by taking values of the parameter of the quartic term ($\lambda$) in between 0 and 1. Authors obtained good results for small values of $\lambda$.
The method of wavefunction expansion over the Oscillator basis is used in ref. \cite{2505.06317} to calculate the energies of the ground and the first six excited states of QAO for a broad range of $\lambda$. Authors improved this expansion method by using the optimised oscillator basis and obtained accurate results. Logarithmic derivation of the eigenfunction approximation(LDoEA) is applied on the Schrodinger equation in ref.\cite{Turbinor} to find the 1st, 2nd and 3rd correction to energy and first correction to wavefunction. Energy eigenvalues for a quantum QAO are calculated with the Dirac operator technique and the Numerov approach in ref. \cite{Adelkun}. Their results which are obtained by two methods diverge for higher excited states.
In ref. \cite{Caffarel}, analytic expressions for the energy and partition functions for the quartic and sextic anharmonic oscillators are proposed.
In ref. \cite{0603165},  quasilinearization method (QLM) is used to find the energies and wave functions of the ground state quartic, and pure quartic oscillators.
In this work, the energy eigen values are calculated for the quadratic ($\frac{g^2 x^2}{2}$), pure quartic ($\lambda x^4 $), and quartic anharmonic oscillator($\frac{g^2 x^2}{2} + \lambda x^4 $) by applying variational method. With $g = 1$ and $\lambda =1/4$, energies are calculated for the ground state and first ten excited states by considering the harmonic oscillator wave functions. For comprehensive study of QAO, energies are calculated for different values of $\lambda$ for ground and first five excited states. The obtained results for the QAO are compared with numerically calculated data reported in \cite{0603165} and observed that energies values are little bit deviate from exact calculated numerical data \cite{0603165}. As energies obtained by variational method depends on the trial wave functions; therefore, a new set of trial wave functions (product of polynomial and exponential terms) is suggested to get more accurate energies for QAO. The computed energy eigen values by this newly suggested wave functions are found to be very close to numerical data. At the end of this work, effect of $\lambda$ on the wave functions is studied.

The rest of the paper is organized as follows. In Sec. 2, theoretical framework for the calculation of the energy eigen values for pure quartic, quadratic, and quartic anharmonic oscillator is presented . Results are reported and discussed in Section 3. Concluding remarks are made in Sec 4.

\section{\textbf{Theoretical Framework} }

Hamiltonian for quartic anharmonic oscillator potential \cite{0603165} can be written as
\begin{equation}
\hat{H}= \frac{\hat{p^2}}{2m} + \frac{g^2 \hat{x}^2}{2} + \lambda \hat{x}^4\\
\label{1}
\end{equation}
when $g = 0$ with nonzero $\lambda$, the potential becomes a “pure quartic” potential. when $\lambda = 0$ with nonzero $g$, the potential becomes “quadratic”.
In above equation, $ \frac{\hat{p^2}}{2m}$ is the kinetic energy and $\frac{g^2 x^2}{2} + \lambda x^4$ is the QAO potential. To calculate the energy, consider one dimensional time independent Schrodinger equation
\begin{equation}
\hat{H} \psi(x) = E \psi(x)\\
\end{equation}

Here $E$ is the energy eigen value, and $\psi$ is the solution (or wave function). For the calculation of energy for QAO, two different set of wave functions are considered:\\
1- Harmonic oscillator wave function (HOWF)\\
2- Product of polynomial and Exponential wave function (PPEWF)

\subsection{Harmonic oscillators wave functions}

Set of wave functions suggested for harmonic oscillators are defined as \cite{zetli}:
\begin{equation}
\psi^{HO}_n(x)= \frac{1}{\sqrt{\sqrt{\pi}2^n n! \alpha}} e^{- x^2 / 2 \alpha^2}H_n(\frac{x}{\alpha}),\\
\label{2}
\end{equation}
where $n=0$ for the ground state, and $n=1,2,3,...$ for the first, 2nd, 3rd,...excited states.
Here $H_n(\frac{x}{\alpha})$ are the Hermite polynomials and defined as
\begin{equation}
H_n(y) = (-1)^n e^{y^2} \frac{d^n}{dy^n}e^{- y^2},
\end{equation}
where $y=\frac{x}{\alpha}$. $\alpha$ is the variational parameter and its value can be found by minimizing the energy
\begin{equation}
\frac{\partial E_n}{\partial \alpha} = 0,
\end{equation}
with
\begin{equation}
E_{n}= \frac{<\psi |\hat{H}|\psi>}{<\psi | \psi>}  \label{3}
\end{equation}

\subsubsection{Energy with Quartic Anharmonic Potential}
Using the natural units $\hbar=1$ and $m=1$, energy expressions obtained for the ground state and first ten excited states of QAO become:

\begin{equation}
E_n=  \frac{2 n+1}{4 \alpha^2} + \frac{2 n+1}{4}\alpha^2 + \frac{3(4[ \sum\limits_{i=0}^n (n-i)]+1) \alpha^4 \lambda }{4}\\ \label{5}
\end{equation}

Considering $\lambda = \frac{1}{4}$ and minimizing the energy expression, $\alpha$ for each energy state is calculated. Substituting the value of $\alpha$ in energy expression, energy values for the ground and ten excited states are calculated. Variational parameter ($\alpha$) and energy are reported in Table 1.

Then $\alpha$ and energies for the ground and first five excited states are calculated for $\lambda = 0, 1/10, 3/10, 1/2, 1, 2, 10, 100, 1000$ to study the effect of $\lambda$ on wavefunctions and reported in Tables (2-8).

\subsubsection{Energy with Pure quartic:}
For pure quartic oscillator, Hamiltonian can be written as
\begin{equation}
\hat{H}= \frac{\hat{p^2}}{2m} + \lambda \hat{x}^4\\ \label{6}
\end{equation}
Considering HOWF as defined above in eq.(\ref{2}) and taking $g=1$, $m=1$, $\lambda=1/4$, following expressions for energy are derived by variational method (explained above in detail in case of QAO).

\begin{equation}
E_n=  \frac{2 n+1}{4 \alpha^2} +  \frac{3(4[ \sum\limits_{i=0}^n (n-i)]+1) \alpha^4}{16}\\ \label{8}
\end{equation}

By minimizing these energy expressions, $\alpha$ for each energy state is calculated and Substituting the value of $\alpha$ in energy expression, energy values for the ground and excited states are calculated. Values of variational parameter ($\alpha$) and energies for pure quartic potential are reported in Table 1.

\subsubsection{Energy with Quadratic Potential:}
For quadratic oscillator, Hamiltonian can be written as
\begin{equation}
\hat{H}= \frac{\hat{p^2}}{2m} + \frac{g^2 \hat{x}^2}{2} \\ \label{9}
\end{equation}
Considering HOWF as defined above in eq.(\ref{2}) and taking $g=1$, $m=1$, $\lambda=1/4$, following expressions for energy are derived by variational method (explained above in detail in case of QAO).
\begin{equation}
E_n=  \frac{2 n+1}{4}(\frac{1}{\alpha^2}+ \alpha^2) \\ \label{10}
\end{equation}

By minimizing these energy expressions, $\alpha$ for each energy state is calculated. Substituting the value of $\alpha$ in energy expression, energy values for the ground and excited states are calculated. Variational parameter ($\alpha$) and energy for quadratic potential are reported in Table 1.

\subsection{Wave function as product of polynomial and exponential}
In this section, a new set of trial wave functions are used for study of QAO. These trial wave functions can be written as:
\begin{equation}
\psi_n(x)= x^n e^{- \acute{\alpha} x^2}(1 - a x + b x^2 - c x^3 + d x^4),\\ \label{11}
\end{equation}
Here, $\acute{\alpha}$, $a$, $b$, $c$, $d$ are the variational parameters. By applying the variational principle for the QAO potential (defined in eq.(\ref{2})), the following energy expressions are obtained for $n= 0,1,2,3,4,5$.
\begin{equation}\label{eq:g2e0}
\resizebox{1.1\hsize}{!}{$
E_0=\frac{(512 \alpha^5 + 192 a^2 \alpha^3 (5 + 2 \alpha) + 768 \alpha^4 (1 + b) +
 480 a \alpha^2 (7 + 2 \alpha) c + 10395 d^2 +
 480 \alpha^3 (4 b + b^2 + 2 d) +
 840 \alpha^2 (2 b^2 + c^2 + 4 d + 2 b d) +
 1890 \alpha (2 c^2 + 4 b d +
    d^2))}{(16 \alpha^2 (4 \alpha (4 \alpha (4 \alpha (a^2 + 4 \alpha) +
         8 \alpha b + 3 b^2) + 24 a \alpha c + 15 c^2) +
   24 \alpha (4 \alpha + 5 b) d + 105 d^2))}$}
\end{equation}

\begin{equation}\label{eq:g2e1}
\resizebox{1.1\hsize}{!}{$
  E_1=\frac{(3 (4 \alpha (4 \alpha (32 \alpha^3 + 20 a^2 \alpha (7 + 2 \alpha) +
           315 b^2 + 80 \alpha^2 (1 + b) + 70 \alpha b (4 + b)) +
        280 a \alpha (9 + 2 \alpha) c + 315 (11 + 2 \alpha) c^2) +
     280 \alpha (8 \alpha^2 + 99 b + 18 \alpha (2 + b)) d +
     3465 (13 +
        2 \alpha) d^2))}{(16 \alpha^2 (4 \alpha (4 \alpha (4 \alpha (3 a^2 \
+ 4 \alpha) + 24 \alpha b + 15 b^2) + 120 a \alpha c + 105 c^2) +
     120 \alpha (4 \alpha + 7 b) d + 945 d^2))}$}
\end{equation}
\begin{equation}\label{eq:g2e2}
\resizebox{1.1\hsize}{!}{$  E_2=\frac{(5 (512 \alpha^5 + 448 a^2 \alpha^3 (9 + 2 \alpha) +
     1792 \alpha^4 (1 + b) + 2016 a \alpha^2 (11 + 2 \alpha) c +
     135135 d^2 + 2016 \alpha^3 (4 b + b^2 + 2 d) +
     5544 \alpha^2 (2 b^2 + c^2 + 4 d + 2 b d) +
     18018 \alpha (2 c^2 + 4 b d +
        d^2)))}{(16 \alpha^2 (4 \alpha (4 \alpha (4 \alpha (5 a^2 +
              4 \alpha) + 40 \alpha b + 35 b^2) + 280 a \alpha c +
        315 c^2) + 280 \alpha (4 \alpha + 9 b) d + 3465 d^2))}$}
\end{equation}
\begin{equation}\label{eq:g2e3}
\resizebox{1.1\hsize}{!}{$  E_3=\frac{(7 (512 \alpha^5 + 576 a^2 \alpha^3 (11 + 2 \alpha) +
     2304 \alpha^4 (1 + b) + 3168 a \alpha^2 (13 + 2 \alpha) c +
     328185 d^2 + 3168 \alpha^3 (4 b + b^2 + 2 d) +
     10296 \alpha^2 (2 b^2 + c^2 + 4 d + 2 b d) +
     38610 \alpha (2 c^2 + 4 b d +
        d^2)))}{(16 \alpha^2 (4 \alpha (4 \alpha (4 \alpha (7 a^2 +
              4 \alpha) + 56 \alpha b + 63 b^2) + 504 a \alpha c +
        693 c^2) + 504 \alpha (4 \alpha + 11 b) d + 9009 d^2))}$}
\end{equation}
\begin{equation}\label{eq:g2e4}
\resizebox{1.1\hsize}{!}{$  E_4=\frac{(9 (512 \alpha^5 + 704 a^2 \alpha^3 (13 + 2 \alpha) +
     2816 \alpha^4 (1 + b) + 4576 a \alpha^2 (15 + 2 \alpha) c +
     692835 d^2 + 4576 \alpha^3 (4 b + b^2 + 2 d) +
     17160 \alpha^2 (2 b^2 + c^2 + 4 d + 2 b d) +
     72930 \alpha (2 c^2 + 4 b d +
        d^2)))}{(16 \alpha^2 (4 \alpha (4 \alpha (4 \alpha (9 a^2 +
              4 \alpha) + 72 \alpha b + 99 b^2) + 792 a \alpha c +
        1287 c^2) + 792 \alpha (4 \alpha + 13 b) d + 19305 d^2))}$}
\end{equation}
\begin{equation}\label{eq:g2e5}
\resizebox{1.1\hsize}{!}{$  E_5=\frac{(11 (512 \alpha^5 + 832 a^2 \alpha^3 (15 + 2 \alpha) +
     3328 \alpha^4 (1 + b) + 6240 a \alpha^2 (17 + 2 \alpha) c +
     1322685 d^2 + 6240 \alpha^3 (4 b + b^2 + 2 d) +
     26520 \alpha^2 (2 b^2 + c^2 + 4 d + 2 b d) +
     125970 \alpha (2 c^2 + 4 b d +
        d^2)))}{(16 \alpha^2 (4 \alpha (4 \alpha (44 a^2 \alpha +
           16 \alpha^2 + 88 \alpha b + 143 b^2) + 1144 a \alpha c +
        2145 c^2) + 1144 \alpha (4 \alpha + 15 b) d + 36465 d^2))}$}
\end{equation}

Minimizing each energy expression, parameters $\acute{\alpha}$, $a$, $b$, $c$, $d$ are found. Values of these variational parameters for $n=0$ are reported in Table 2; while energies are reported in Table 3. For excited states ($n=1,2,3,4,5$), Energies along with variational parameters are reported in Tables (4-8).

\section{3-\textbf{Results and Discussion}}
\noindent
Eigen energies, calculated with wave functions of simple harmonic oscillator for quadratic, quartic and pure quartic oscillators, are reported in Table 1 for the ground and first ten excited states by taking $g=1$ and $\lambda=\frac{1}{4}$. Comparing the results, it is observed that energies of QAO are greater than the quadratic and pure quartic potentials. It is also observed that energies are increasing towards higher excited states. For different values of $\lambda$ and $g=1$, the energies of quartic potential for ground state and first five excited states are reported in Tables (3-8). In Table 3, the energies for the ground state are reported and compared with numerically calculated (exact) data \cite{0603165}. It is observed that our calculated results with HOWF for QAO are greater than the exact values. Maximum $\%$ errors between energies calculated by WKB \cite{0603165}, QLM \cite{0603165} and expansion method \cite{2505.06317} are found to be 18.1450, 9.0 and 8.2 respectively; while the maximum $\%$ error between exact values and our calculated values is found to be 1.9977. This proves that our method with the selected wave functions gives better results than WKB method \cite{0603165}, QLM\cite{0603165} and expansion method \cite{2505.06317}.

To achieve more accurate energies, trial wave functions are considered as product of polynomial and exponential functions. With this newly suggested set of wave functions, our calculated energies come close to numerical data (exact) \cite{0603165}. For ground state (n=0), maximum $\%$ error in energy with newly suggested wave functions is equal to 0.5610.

Ground-state and first excited state wave functions of QAO potential at $\lambda = 0,1,10,100,1000$ are shown in Fig.1 and 2 respectively. In part (a) of the figures, wave functions for SHO are shown while the newly suggested wave functions (polynomial plus exponential) are shown in part (b). From Figures, it is observed that both forms of wave functions  (SHO or PPE) have the similar behaviour i.e; wave functions become narrower and sharper when we increase the value of $\lambda$. In other words, its concluded that the curves for wave functions are converging toward higher value of $\lambda$. Height of the WFs increases toward the higher value of $\lambda$. It is noted that number of zero crossing nodes in each curve are equal to ``n"  which is an expected result for both the harmonic and anharmonic oscillator \cite{2505.06317,0603165}. For the ground state, peak lies at the origin for each value of $\lambda$. But for the excited state wave functions, peak shifts toward the origin with the increase of $\lambda$.

\begin{table}[h!]
\centering
\caption{Variational parameter $\alpha$ and energy values for quadratic, quartic anharmonic and quartic potential for ground and first ten excited states by taking $g=1$ and $\lambda = \frac{1}{4}.$}
\begin{tabular}{|c| c| c| c| c| c| c|}
\hline\hline
\hspace{0.3cm} State \hspace{0.3cm} & \hspace{0.3cm} $\alpha_{V \propto \frac{x^2}{2}}$ \hspace{0.3cm}& \hspace{0.3cm} $E_{V \propto \frac{x^2}{2}}$ \hspace{0.3cm}& \hspace{0.3cm} $\alpha_{V \propto \frac{x^2}{2}+\frac{x^4}{4}}$ \hspace{0.3cm}& \hspace{0.3cm} $E_{V \propto \frac{x^2}{2}+\frac{x^4}{4}}$ \hspace{0.3cm}& \hspace{0.3cm} $\alpha_{V \propto \frac{x^4}{4}}$ \hspace{0.3cm}& \hspace{0.3cm} $E_{V \propto \frac{x^4}{4}}$ \hspace{0.3cm}\\
\hline
0     &1      &0.5       &0.835913        &0.624016      &0.934655      &0.429268    \\
1     &1      &1.5       &0.790422        &2.03496       &0.858374      &1.52686     \\
2     &1      &2.5       &0.74854         &3.69654       &0.797057      &2.95136     \\
3     &1      &3.5       &0.718034        &5.54254       &0.755981      &4.59311      \\
4     &1      &4.5       &0.694574        &7.53854       &0.72593       &6.40449     \\
5     &1      &5.5       &0.675683        &9.66296       &0.702525      &8.35796     \\
6     &1      &6.5       &0.659954        &11.9008       &0.683501      &10.4351     \\
7     &1      &7.5       &0.646527        &14.2408       &0.667555      &12.6226      \\
8     &1      &8.5       &0.634846        &16.6742       &0.65388       &14.9102     \\
9     &1      &9.5       &0.624529        &19.1939       &0.641944      &17.2898     \\
10    &1      &10.5      &0.615306        &21.7941       &0.631378      &19.7548     \\

  \hline\hline
\end{tabular}
\label{quark-model-parameters}
\end{table}

\begin{table}[h!]
\scriptsize
\setlength{\tabcolsep}{3pt}
\centering
\caption{Variational parameters for ground state ($n=0$) of QAO at different values of $\lambda$ by taking $g=1$.}
\begin{tabular}{|c| c| c| c |c| c| c|}
\hline\hline
\hspace{0.1cm} $\lambda$ \hspace{0.2cm} & \hspace{0.2cm} $\alpha$ \hspace{0.2cm}& \hspace{0.2cm} $\alpha'$ \hspace{0.2cm}&\hspace{0.2cm} $a$ \hspace{0.2cm}& \hspace{0.2cm} $b$ \hspace{0.2cm}& \hspace{0.2cm} $c$ \hspace{0.2cm}& \hspace{0.2cm} $d$ \hspace{0.2cm} \\
\hline
0       & 1.0000     & 0.5159 & $-6.3812\times 10^{-8}$ & 0.0158 & $4.5172\times 10^{-9}$ & 0.0001 \\
1/10    & 0.9049     & 0.9593 & $1.5792\times 10^{-8}$  & 0.3970 & $-6.2552\times 10^{-9}$ & 0.0650 \\
3/10    & 0.8201     & 1.2758 & $1.1366\times 10^{-8}$  & 0.6298 & $-3.2124\times 10^{-8}$ & 0.1704 \\
1/2     & 0.7734     & 1.4768 & $-2.6587\times 10^{-8}$ & 0.7695 & $1.1979\times 10^{-8}$  & 0.2579 \\
1       & 0.7071     & 1.8188 & $-2.7498\times 10^{-8}$ & 0.9985 & $2.0280\times 10^{-8}$  & 0.4403 \\
2       & 0.6451     & 1.4587 & $-2.1643\times 10^{-8}$ & 0.5405 & $1.8083\times 10^{-9}$  & -0.1850 \\
10      & 0.5000     & 2.4254 & $-2.7190\times 10^{-8}$ & 0.9866 & $-3.5926\times 10^{-8}$ & -0.5607 \\
100     & 0.3435     & 5.1645 & $-2.4426\times 10^{-7}$ & 2.1850 & $9.8765\times 10^{-7}$  & -2.6425 \\
1000    & 0.2345     & 11.0987& $1.7278\times 10^{-7}$  & 4.7358 & $-2.2411\times 10^{-6}$ & -12.3060 \\
  \hline\hline
\end{tabular}
\label{quark-model-parameters}
\end{table}

\begin{table}[h!]
\scriptsize
\setlength{\tabcolsep}{2pt}
\centering
\caption{Comparison of our calculated energy values of ground state QAO with others calculated results at different values of $\lambda$.}
\begin{tabular}{|c| c| c|c| c| c| c| c |c|c|c|c|}
\hline\hline
$\lambda$  &  $Exact$ &  $E_{WKB}$ &  $E_{QLM}$ & $E$ & $E_{SHOWF}$ & $E_{PPEWF}$ & $\% Err $ & $\%Err $ & $\% Err $ &  $\% Err $ &  $\% Err $ \\
 & \cite{0603165} & \cite{0603165} & \cite{0603165} & \cite{2505.06317} & This work & This work & $E_{WKB}$ & $ E_{QLM}$ & $E$ \cite{2505.06317} & $ E_{SHOWF}$ & $E_{PPEWF}$ \\
 \hline
0     &0.5000    &0.5000     &0.5000   &--     &0.5000   &0.5000    &0         &0       &--      &0        &0       \\
1/10  &0.5592    &0.5333     &0.5615   &0.5591 &0.5603   &0.5591    &4.6267    &0.4185  &0.0179  &0.2069   &0.0179   \\
3/10  &0.6380    &0.5847     &0.6471   &0.6903 &0.6416   &0.6380    &8.3591    &1.4201  &8.2003  &0.5705   &0.0000    \\
1/2   &0.6962    &0.6254     &0.7113   &0.6962 &0.7016   &0.6962    &10.1698   &2.1661  &0.0000  &0.7874   &0.0000    \\
1     &0.8038    &0.7042     &0.8309   &0.8037 &0.8125   &0.8038    &12.3879   &3.3753  &0.0124  &1.0861   &0.0000    \\
2     &0.9516    &0.8167     &0.9958   &--     &0.9644   &0.9517    &14.1766   &4.6450  &--      &1.3487   &0.0105    \\
10    &1.5050    &1.2541     &1.6109   &1.5049 &1.5313   &1.5053    &16.6681   &7.0354  &0.0066  &1.7462   &0.0199    \\
100   &3.1314    &2.5718     &3.4004   &--     &3.1924   &3.1321    &17.8698   &8.5908  &--      &1.9499   &0.0223    \\
1000  &6.6942    &5.4795     &7.2974   &--     &6.8279   &6.6566    &18.1451   &9.0111  &--      &1.9977   &0.5610    \\
\hline\hline
\end{tabular}
\label{quark-model-parameters}
\end{table}

\begin{table}[h!]
\scriptsize
\setlength{\tabcolsep}{3pt}
\centering
\caption{Results for parameters and energies of QAO for 1st excited state ($n=1$) at different values of $\lambda$.}
\begin{tabular}{|c|c|c|c|c|c|c|c|c|c|}
\hline\hline
$\lambda$ & $\alpha$ & $\alpha'$ & $a$ & $b$ & $c$ & $d$ & $E_{V1}$ & $E_{V2}$ &  $E \cite{2505.06317}$\\
\hline
0      & --      & 0.8650 & $-0.1233$ & 0.3005 & 0.0089   & 0.1451    & --       & 1.5072   & ---- \\
1/10   & 0.8688  & 0.9910 & $0.0000$  & 0.3922 & $0.0000$ & 0.0705    & 1.7734   & 1.7695  & 1.7695 \\
3/10   & 0.7734  & 1.3296 & $0.0000$  & 0.6099 & $0.0000$ & 0.1820    & 2.1050   & 2.0947  & 2.0946 \\
1/2    & 0.7247  & 1.5437 & $0.0000$  & 0.7392 & $0.0000$ & 0.2736    & 2.3391   & 2.3245  & 2.3244 \\
1      & 0.6581  & 1.9068 & $0.0000$  & 0.9510 & $0.0000$ & 0.4634    & 2.7599   & 2.7380  & 2.7379 \\
2      & 0.5937  & 1.6078 & $0.0000$  & 0.5749 & $0.0000$ & -0.1762   & 3.3240   & 3.2932  & 3.2929 \\
10     & 0.4606  & 2.6879 & $0.0000$  & 1.0393 & $0.0000$ & -0.5321   & 5.3821   & 5.3223  & 5.3216 \\
100    & 0.3157  & 5.7370 & $0.0000$  & 2.2918 & $0.0000$ & -2.5034   & 11.3249  & 11.1888 & 11.1873 \\
1000   & 0.2154  & 12.3351& $0.0000$  & 4.9626 & $0.0000$ & -11.6542  & 24.2722  & 23.9756 & 23.9722 \\
\hline
\hline\hline
\end{tabular}
\label{quark-model-parameters}
\end{table}

\begin{table}[h!]
\scriptsize
\setlength{\tabcolsep}{2pt}
\centering
\caption{Results for parameters and energies of QAO for 2nd excited state ($n=2$) at different values of $\lambda$.}
\begin{tabular}{|c|c|c|c|c|c|c|c|c|c|c|}
\hline\hline
$\lambda$ & $\alpha$ & $\alpha'$ & $a$ & $b$ & $c$ & $d$ & $EV1$ & $EV2$ & $E \cite{2505.06317}$ \\
\hline
  0    & ---    & 1.1095  & $3.1969\times 10^{-8}$  & -0.3322 & $-9.6906\times 10^{-9}$ & 0.2092   & ----    & 1.6234  & ----   \\
1/10   & 0.8326 & 1.5141  & $1.8717\times 10^{-8}$  & -0.4210 & $1.9628\times 10^{-9}$  & 0.4114   & 3.1382  & 1.9148  & 3.1386  \\
3/10   & 0.7311 & 1.9328  & $5.3775\times 10^{-8}$  & -0.5188 & $-3.7516\times 10^{-9}$ & 0.6886   & 3.8424  & 2.2679  & 3.8448  \\
1/2    & 0.6819 & 2.2106  & $-7.4693\times 10^{-9}$ & -0.5854 & $3.2488\times 10^{-8}$  & 0.9103   & 4.3235  & 2.5175  & 4.3275   \\
1      & 0.6164 & 2.6917  & $9.2245\times 10^{-8}$  & -0.7026 & $-1.6350\times 10^{-7}$ & 1.3649   & 5.1724  & 2.9667  & 5.1793   \\
2      & 0.5544 & 3.3171  & $9.6515\times 10^{-8}$  & -0.8570 & $-6.3216\times 10^{-8}$ & 2.0898   & 6.2933  & 3.5693  & 6.3038    \\
10     & 0.4285 & 5.5301  & $2.5052\times 10^{-7}$  & -1.4107 & $-4.9787\times 10^{-7}$ & 5.8680   & 10.3244 & 5.7714  & 10.3405   \\
100    & 0.2933 & 11.7895 & $-3.1944\times 10^{-7}$ & -2.9905 & $4.2280\times 10^{-6}$  & 26.7911  & 21.8535 & 12.1359 & 21.9068   \\
1000   & 0.2000 & 25.3418 & $-1.4508\times 10^{-7}$ & -6.4201 & $-7.2534\times 10^{-6}$ & 123.9130 & 46.9000 & 26.0065 & 47.0173   \\
\hline\hline
\end{tabular}
\label{quark-model-parameters-n2}
\end{table}
\begin{table}[h!]
\scriptsize
\setlength{\tabcolsep}{2pt}
\centering
\caption{Results for parameters and energies of QAO for 3rd excited state ($n=3$) at different values of $\lambda$.}
\begin{tabular}{|c|c|c|c|c|c|c|c|c|c|}
\hline\hline
$\lambda$ & $\alpha$ & $\alpha'$ & $a$ & $b$ & $c$ & $d$ & $EV1$ & $EV2$ & $E \cite{2505.06317}$ \\
\hline
0      & ---- & 1.3346 & $1.8644\times 10^{-8}$ & $-0.5249$ & $-1.2182\times 10^{-8}$ & 1.1095    & --- & 1.7791 & --- \\
1/10   & 0.8045 & 1.8015 & $4.5514\times 10^{-8}$ & $-0.6932$ & $-3.7409\times 10^{-9}$ & 0.4299  & 4.6219 & 2.1045 & 4.6288 \\
3/10   & 0.7005 & 2.2901 & $1.7142\times 10^{-8}$ & $-0.8722$ & $6.1995\times 10^{-9}$  & 0.7042  & 5.7795 & 2.4976 & 5.7966 \\
1/2    & 0.6516 & 2.6153 & $8.1309\times 10^{-9}$ & $-0.9922$ & $-1.8409\times 10^{-8}$ & 0.9232  & 6.5548 & 2.7749 & 6.5784 \\
1      & 0.5875 & 3.1797 & $2.9808\times 10^{-8}$ & $-1.2013$ & $-4.8260\times 10^{-8}$ & 1.3725  & 7.9079 & 3.2734 & 7.9424 \\
2      & 0.5274 & 3.9144 & $2.1630\times 10^{-8}$ & $-1.4745$ & $-3.0699\times 10^{-8}$ & 2.0886  & 9.6796 & 3.9416 & 9.7273\\
10     & 0.4069 & 6.5176 & $-1.2841\times 10^{-7}$& $-2.4461$ & $-3.1815\times 10^{-7}$ & 5.8204  & 15.9993 & 6.3803 & 16.0902  \\
100    & 0.2782 & 14.3617& $-4.9823\times 10^{-2}$& $-5.6885$ & $-1.5536\times 10^{-2}$ & 29.4760 & 33.9779 & 13.0230 & 34.1825 \\
1000   & 0.1897 & 29.8474& $-4.0516\times 10^{-7}$& $-11.1800$& $7.7075\times 10^{-6}$  & 122.4120& 72.9741 & 28.7684 & 73.4191 \\
\hline\hline
\end{tabular}
\label{quark-model-parameters}
\end{table}

\begin{table}[h!]
\scriptsize
\setlength{\tabcolsep}{2pt}
\centering
\caption{Results for parameters and energies of QAO for 4th excited state ($n=4$) at different values of $\lambda$.}
\begin{tabular}{|c|c|c|c|c|c|c|c|c|c|}
\hline
$\lambda$ & $\alpha$ & $\alpha'$ & $a$ & $b$ & $c$ & $d$ & $EV1$ & $EV2$ & $E \cite{2505.06317}$ \\
\hline
0     & ---    & 1.5112 & $-3.5220\times10^{-8}$ & -0.6034 & $-7.4813\times10^{-10}$    & 0.2186  & ---     & 1.9261 & --- \\
1/10  & 0.7821 & 2.0387 & $2.3034\times10^{-8}$  & -0.8050 & $-1.9464\times10^{-8}$  & 0.4048  &6.2052   & 2.2870    & 6.2203 \\
3/10  & 0.6771 & 2.5905 & $-1.0023\times10^{-8}$ & -1.0176 & $9.3589\times10^{-10}$   & 0.6592  & 7.8782  & 2.7206   & 7.9118 \\
1/2   & 0.6287 & 2.9579 & $3.7213\times10^{-8}$  & -1.1597 & $-2.1960\times10^{-8}$   & 0.8623  & 8.9838  & 3.0258   & 9.0286 \\
1     & 0.5659 & 3.5956 & $-3.8471\times10^{-8}$ & -1.4067 & $3.9776\times10^{-8}$    & 1.2788  & 10.9000 & 3.5735   & 10.9636 \\
2     & 0.5075 & 4.4257 & $-1.5616\times10^{-8}$ & -1.7290 & $-1.4541\times10^{-8}$   & 1.9424  & 13.3951 & 4.3067   & 13.4813 \\
10    & 0.3910 & 7.3677 & $-5.2779\times10^{-8}$ & -2.8730 & $1.3048\times10^{-7}$    & 5.4005  & 22.2484 & 6.9795   & 22.4088 \\
100   & 0.2672 & 11.8257& $52.4496$              & 5.7943  & $-7.2998$                & -5.1408 & 47.3495 & 18.9854  & 47.70725 \\
1000  & 0.1822 & 26.9273& $12.0423$              & 25.9020 & $1.6869$                 & 4.8171  & 101.7400 & 40.5811 & 102.514 \\
\hline \hline
\end{tabular}
\end{table}
\begin{table}[h!]
\scriptsize
\setlength{\tabcolsep}{2pt}
\centering
\caption{Results for parameters and energies of QAO for 5th excited state ($n=5$) at different values of $\lambda$.}
\begin{tabular}{|c| c| c| c| c| c| c| c| c| c|}
\hline\hline
$\lambda$ & $\alpha$ & $\alpha'$ & $a$ & $b$ & $c$ & $d$ & $E V1$ & $E V2$ & $E \cite{2505.06317}$ \\
\hline
0      & ---     & 1.6636  & $2.8055\times 10^{-8}$  & $-0.6352$  & $-1.3414\times 10^{-8}$  & 0.2017  & --      & 2.0635 & --- \\
1/10   & 0.7637  & 2.2501  & $2.5363\times 10^{-8}$  & $-0.8532$  & $-1.3567\times 10^{-8}$  & 0.3736  & 7.8752  & 2.4601 & 7.8998 \\
3/10   & 0.6583  & 2.8615  & $-4.6129\times 10^{-8}$ & $-1.0817$  & $1.8465\times 10^{-8}$   & 0.6078  & 10.1151 & 2.9335 & 10.1665 \\
1/2    & 0.6105  & 3.2682  & $1.1034\times 10^{-8}$  & $-1.2340$  & $-1.2200\times 10^{-8}$  & 0.7947  & 11.5810 & 3.2659 & 11.6987 \\
1      & 0.5488  & 3.9737  & $1.7574\times 10^{-8}$  & $-1.4984$  & $6.3936\times 10^{-9}$   & 1.1777  & 14.1090 & 3.8616 & 14.2031 \\
2      & 0.4918  & 4.8919  & $4.3208\times 10^{-8}$  & $-1.8430$  & $-4.9671\times 10^{-8}$  & 1.7880  & 17.3877 & 4.6580 & 17.5141 \\
10     & 0.3786  & 8.1444  & $-7.4142\times 10^{-8}$ & $-3.0649$  & $8.4135\times 10^{-8}$   & 4.9668  & 28.9793 & 7.5573 & 29.2115 \\
100    & 0.2586  & 13.2819 & 1.7275                  & $-3.7438$  & $-2.1893$                & 5.0732  & 61.7660 & 20.4899 & 62.2812 \\
1000   & 0.1763  & 28.5900 & 480.8010                & 4.9273     & 3.0109                   & 31.1223 & 132.7600 & 44.8562 & 133.8769 \\
\hline\hline
\end{tabular}
\label{quark-model-parameters}
\end{table}

\begin{figure}[h!]
\centering
\begin{subfigure}[b]{0.49\textwidth}
\centering
\includegraphics[width=\textwidth]{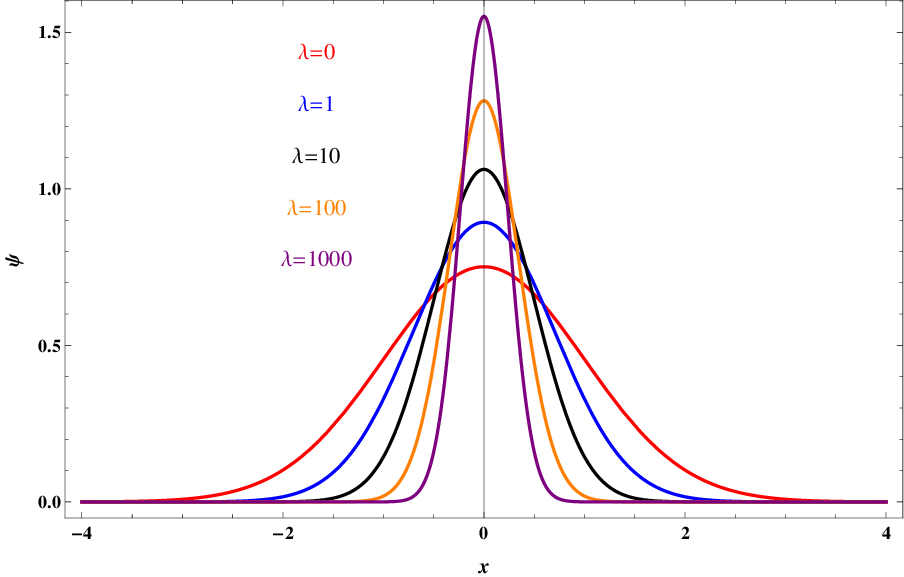}
\caption{}
\label{fig:a}
\end{subfigure}
\hfill
\begin{subfigure}[b]{0.49\textwidth}
\centering
\includegraphics[width=\textwidth]{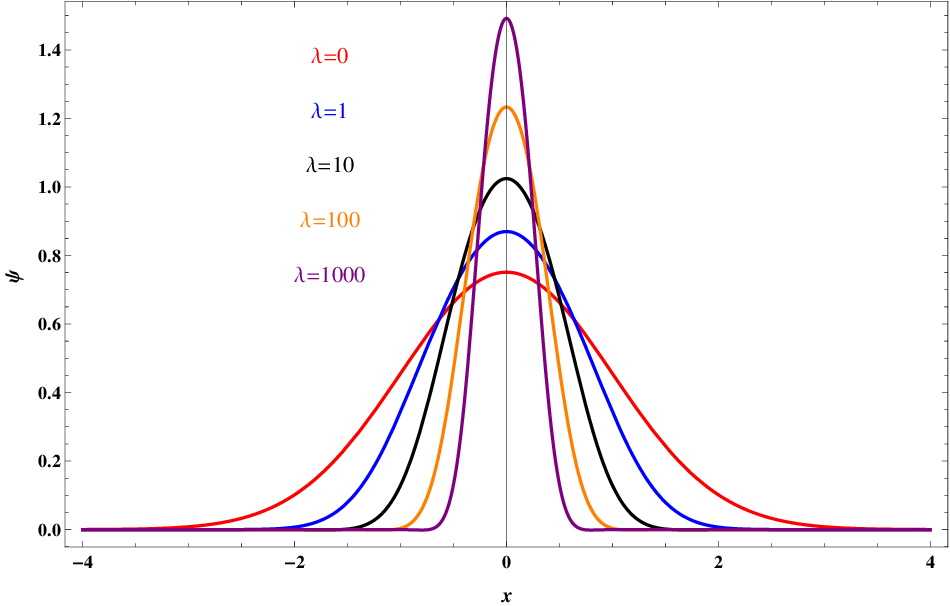}
\caption{}
\label{fig:b}
\end{subfigure}
\caption{Wave functions of QAO for the ground state at different values of $\lambda$ (a) SHO WF  (b) product of polynomial and exponential function.}
\label{fig:combined}
\end{figure}

\begin{figure}[h!]
\centering
\begin{subfigure}[b]{0.49\textwidth}
\centering
\includegraphics[width=\textwidth]{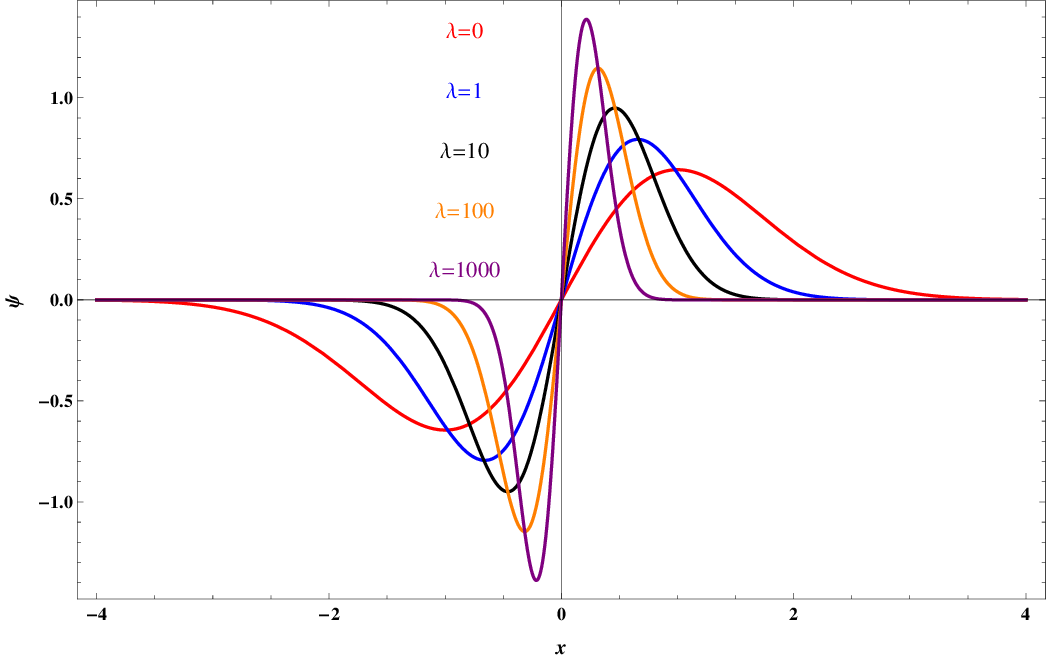}
\caption{}
\label{fig:a}
\end{subfigure}
\hfill
\begin{subfigure}[b]{0.49\textwidth}
\centering
\includegraphics[width=\textwidth]{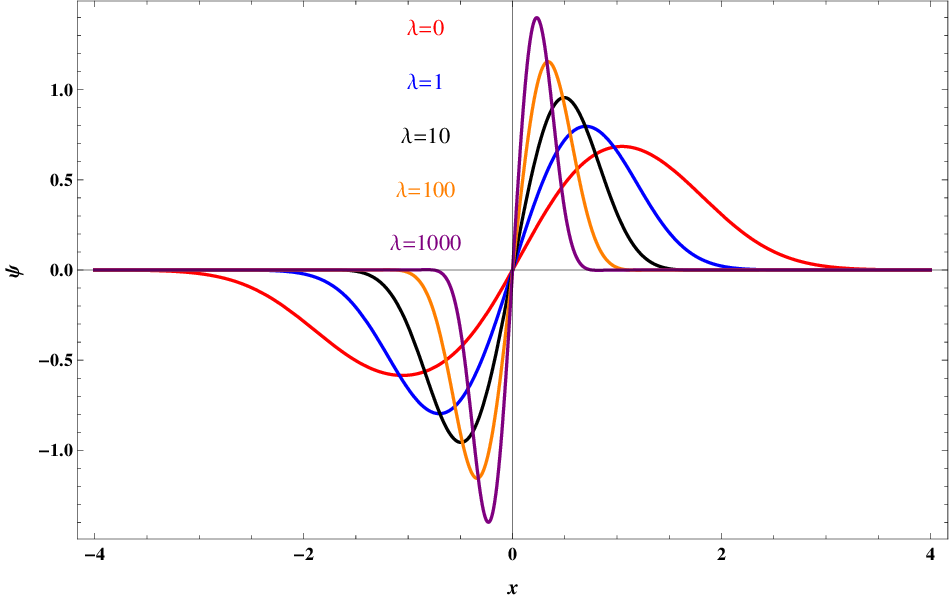}
\caption{}
\label{fig:b}
\end{subfigure}
\caption{Wave functions for first excited state of QAO at different values of $\lambda$ (a) SHO WF  (b) product of polynomial and exponential function.}
\label{fig:combined}
\end{figure}

\section{Conclusion}
As the main purpose of this work is to investigate the physical models through analytical solutions which are very important to understand the hidden structures of physical systems. The variational method can successfully applied to solve many other similar problems as well. Comparison of our results for energies with the exact numerically calculated energies proves that our suggested trial wave functions give more accurate results than the results calculated by WKB, QLM, and wave function expansion method reported in Refs.\cite{0603165,2505.06317}. These wave functions can be used to find the physical properties of quartic and pure quartic oscillators.

\end{document}